\documentclass[aps,showpacs,12pt]{revtex4}
\usepackage{amsfonts,amsmath,amsthm}
\usepackage{graphicx}
\usepackage{eucal}

\begin{document}

\title{Theoretical investigation of finite size effects at DNA melting}
\author{Sahin BUYUKDAGLI and Marc $\mbox{JOYEUX}^{(\sharp)}$}
\affiliation{Laboratoire de Spectrom\'{e}trie Physique (CNRS UMR
5588), Universit\'{e} Joseph Fourier - Grenoble 1, BP 87, 38402 St
Martin d'H\`{e}res, FRANCE }

\begin{abstract}
We investigated how the finiteness of the length of the sequence
affects the phase transition that takes place at DNA melting
temperature. For this purpose, we modified the Transfer Integral
method to adapt it to the calculation of both extensive (partition
function, entropy, specific heat, etc) and non-extensive (order
parameter and correlation length) thermodynamic quantities of finite
sequences with open boundary conditions, and applied the modified
procedure to two different dynamical models. We showed that rounding
of the transition clearly takes place when the length of the
sequence is decreased. We also performed a finite-size scaling
analysis of the two models and showed that the singular part of the
free energy can indeed be expressed in terms of an homogeneous
function. However, both the correlation length $\xi$  and the
average separation between paired bases $\left<y\right>$ diverge at
the melting transition, so that it is no longer clear to which of
these two quantities the length $L$ of the system should be
compared. Moreover, Josephson's identity is satisfied for none of
the investigated models, so that the derivation of the
characteristic exponents which appear, for example, in the
expression of the specific heat, requires some care.

\vspace{1cm} $^{(\sharp)}$email : Marc.JOYEUX@ujf-grenoble.fr
\end{abstract}

\maketitle
\newpage

\section{Introduction}

Real systems manifesting critical behavior have necessarily finite
volume. However it is well-known that the finiteness of the system
size lets the critical singularities disappear and smears out the
phase transition (see for example [1-5]). The obvious argument which
may reconcile these two aspects is that, as the finite size of the
system is increased and passes through a critical value which
characterizes the border of the thermodynamical domain, the
thermodynamical limit is approximately reached and consequently the
critical singularities manifest themselves. Thus it is crucial to
thoroughly understand the evolution of singularities with respect to
the system size and estimate the critical size above which the
thermodynamical limit is attained.

An efficient tool to analyze the volume dependence of a critical
phenomenon is the \textit{finite size scaling theory}. Besides
providing information on the rounding of critical singularities and
the shift in the critical point, this theory is also an alternative
way to determine critical exponents characterizing the phase
transition.

Finite size scaling theory was developed by Fisher and Barber in the
early seventies \cite{6}. During the last thirty years it has been
applied to various systems exhibiting both first and second order
phase transitions. Among hundreds of subjects we can mention the
study of finite size effects at first order transitions by gaussian
approximation [7-8],  the Gibbs ensemble \cite{9}, five dimensional
Ising model [10-11], percolation models [12-13], stochastic
sandpiles \cite{14}, six-dimensional Ising system \cite{15},
Baxter-Wu model \cite{16}, two dimensional anisotropic Heisenberg
model \cite{17}...

In nature, the majority of phase transitions are sharp and
discontinuous first order transitions. Experimental UV absorption
spectra of diluted DNA solutions reveal that the DNA melting
transition belongs to this class. A widely used dynamical non-linear
DNA model was proposed twenty years ago by Peyrard and Bishop
\cite{18}. This model involving harmonic interaction terms between
successive base pairs was later improved by the contribution of
Dauxois \cite{19} and the new model (DPB model) yields a sharp
transition. We recently proposed an alternative DNA model (JB model)
which respects the finiteness of the stacking energy, and showed
that this model also exhibits a sharp first order phase transition
\cite{20}. Then in \cite{21} we showed that for both models the
generalized homogeneity assumption is not respected so that
Josephson's identity (also known as the hyperscaling relation) is
not valid. We tentatively explained this fact by the divergence of
the order parameter at the critical point.

The goal of this article is to investigate the sequence length
dependence of the DNA melting transition. This is an important point
since experiments dealing with DNA molecules are carried out with
various sequence lengths. To this end we employ a modified transfer
integral method adapted to finite chains with open boundary
conditions. The two hamiltonian DNA models to be studied, i.e. the
DPB and the JB models, are briefly described in Sec. II. Section III
deals with the transfer matrix theory for finite linear chains and
Sec. IV is devoted to the finite size scaling analysis which leads
to a better understanding of finite size effects.

\section{Non-linear Hamiltonian models for DNA}

The Hamiltonians of the two DNA models whose critical behaviour is
studied in this article are of the form

\begin{equation}\label{0}
H=\sum_{n=1}^N\left\{\frac{p_n^2}{2m}+V_M(y_n)+W(y_n,y_{n-1})\right\}
\end{equation}
where $y_n$ is the transverse stretching of the hydrogen bond
between the nth pair of bases, while the one-particle Morse
potential term

\begin{equation}\label{0.1}
V_M(y_n)=D\left(1-e^{-ay_n}\right)^2
\end{equation}
models the binding energy of the same hydrogen bond. The choice of
the nearest-neighbor interaction potential $W(y_n,y_{n-1})$ is
crucial since the type of the transition, which is a collective
effect, depends primarily on its form. The DPB model \cite{19}
assumes that the stacking energy is of the form

\begin{equation}\label{0.2}
W(y_n,y_{n-1})=\frac{K}{2}(y_n-y_{n-1})^2\left[1+\rho
e^{-\alpha(y_n+y_{n-1})}\right]\hspace{0.5mm}.
\end{equation}
This non-linear stacking interaction has the particularity of having
a coupling constant which drops from $K(1+\rho)$ to $K$ as the
critical point is approached. This decreases the rigidity of the DNA
chain  close to the dissociation and yields a sharp, first-order
transition.

This interaction potential still has the inconvenience that the
stacking energy diverges when two paired bases separate. Taking into
account the finiteness of the interaction between adjacent bases, we
proposed a potential of the form \cite{20}

\begin{equation}\label{0.3}
W(y_n,y_{n-1})=\frac{\Delta
H}{2}\left[1-e^{-b(y_n-y_{n-1})^2}\right]+K_b(y_n-y_{n-1})^2
\end{equation}
which, contrary to the model (\ref{0.2}), depends only on the
distance between base pairs. The small harmonic term, whose constant
$K_b$ is 2000 times smaller than the parameter $K$ of the DPB model,
was introduced in order to take into account the stiffness of the
phosphate-sugar backbone.

Numerical values of the parameters are those of Refs. [19,20], that
is $D=0.03\hspace{0.5mm}eV$, $a=4.5\hspace{0.5mm}\mathrm{\r A}^{-1}$,
$\alpha=0.35\hspace{0.5mm}\mathrm{\r A}^{-1}$,
$K=0.06\hspace{0.5mm}eV\hspace{0.5mm}\mathrm{\r A}^{-2}$, $\rho=1$ for the
DPB model, and $D=0.04\hspace{0.5mm}eV$,
$a=4.45\hspace{0.5mm}\mathrm{\r A}^{-1}$, $\Delta H=0.44\hspace{0.5mm}eV$,
$K_b=10^{-5}\hspace{0.5mm}eV\hspace{0.5mm}\mathrm{\r A}^{-2}$ and
$b=0.10\hspace{0.5mm}\mathrm{\r A}^{-2}$ for the JB model.

\section{ Transfer integral method for finite chains with open boundary conditions}

\subsection{The partition function and extensive thermodynamic quantitites}

Let us define the Transfer Integral (TI) kernel according to

\begin{equation}\label{1}
K(y_{n},y_{n-1}) = exp\left[-\beta
\left\{V_M(y_{n})/2+V_M(y_{n-1})/2+W(y_n,y_{n-1})\right\} \right]
\end{equation}
where $\beta=(k_BT)^{-1}$ is the inverse temperature. In the
following analysis we deal only with sequences having open boundary
conditions. Then the partition function of the system can be
expressed as

\begin{equation}\label{2}
Z = \int \,dy_1\,dy_2 \cdot\cdot\cdot dy_N e^{-\beta V_M(y_1)/2}
K\left(y_2,y_1\right) K\left(y_3,y_2\right) \cdot\cdot\cdot
K\left(y_N,y_{N-1}\right) e^{-\beta V_M(y_N)/2}.
\end{equation}
The TI method consists in expanding the kernel of Eq. (\ref{1}) in
an orthonormal basis

\begin{equation}\label{3}
K\left(y_n,y_{n-1}\right) = \sum_{i}\lambda_i
\Phi_i(y_n)\Phi_i(y_{n-1})
\end{equation}
where the $\{\Phi_i\}$ and $\{\lambda_i\}$ are the eigenvalues and
eigenvectors of the integral operator and satisfy

\begin{equation}
\int
\,dx\hspace{1mm}K(x,y)\hspace{0.5mm}\Phi_i(x)=\lambda_i\Phi_i(y)\hspace{0.5mm}.
\end{equation}
This integral equation was solved by diagonalizing the symmetric TI
operator $K(x,y)$ on a regularly spaced grid defined between
$y_{min}=-200/a$ and $y_{max}=4000/a$  with $1/a$ intervals.
Numerical integrations were performed on the same grid.

In this study we extended the transfer matrix approach for open
chains developed in \cite{22} to adapt it to the calculation of the
order parameter $\left<y\right>$ and the correlation length $\xi$.
Let us first consider extensive thermodynamic quantities. By
introducing

\begin{equation}\label{4}
a_i = \int \,dy e^{-\beta V_M(y)/2}\Phi_i(y)
\end{equation}
and by substituting the kernel expansion of Eq. (\ref{3}) into Eq.
(\ref{2}), we get

\begin{equation}\label{5}
Z = \sum_{i}a_i^2\lambda_i^{N-1}.
\end{equation}
Determination of the partition function then allows the computation
of extensive quantities  of the system such as the free energy, the
entropy and the specific heat :

\begin{equation}\label{5.5}
\begin{split}
F&=-k_BT\ln(Z)\\
S&=-\frac{\partial F}{\partial T}\\
C_V&=-T\frac{\partial^2F}{\partial T^2}\hspace{1 mm}.
\end{split}
\end{equation}

In the thermodynamical limit $N\rightarrow\infty$ the major
contribution to the partition function arises from the largest
eigenvalue $\lambda_1$ and in this limit it is reasonable to drop
the eigenvalues with $i\geq 2$. Neverthless, we will consider large
DNA molecules as well as small ones. Consequently as many
eigenvalues as possible must be taken into account in numerical
computations. From the practical point of view, it was found that
considering the first 400 eigenvalues is enough to insure numerical
convergence of the results presented below.
\subsection{The order parameter and the correlation length}

The order parameter of DNA melting transition is the mean separation
of the bases averaged over the sites of the sequence :

\begin{equation}\label{9}
\left<y\right> = \frac{1}{N}\sum_{n=1}^N \left<y_n\right>.
\end{equation}
In order to reduce $\left<y_n\right>$ to a form depending only on
the eigenvalues and eigenvectors of the TI operator, we first write
it as

\begin{equation}\label{10}
\left<y_n\right> = \frac{1}{Z}\int \,dy_1\,dy_2 \cdot\cdot\cdot dy_N
\hspace{0.5 mm}y_n\hspace{0.5 mm}e^{-\beta V_M(y_1)/2}
K\left(y_2,y_1\right) K\left(y_3,y_2\right) \cdot\cdot\cdot
K\left(y_N,y_{N-1}\right) e^{-\beta V_M(y_N)/2}.
\end{equation}
Substituting Eq. (\ref{3}) into Eq. (\ref{10}) and defining

\begin{equation}\label{11}
\begin{split}
b_i &= \int \,dy e^{-\beta V_M(y)/2}\Phi_i(y)\hspace{0.5 mm}y
\hspace{0.5 mm}\\
Y_{ij}^{(1)} &= \int \,dy \Phi_i(y)\hspace{0.5 mm}y\hspace{0.5
mm}\Phi_j(y)\hspace{0.5 mm},
\end{split}
\end{equation}
we obtain

\begin{equation}\label{13}
\left<y_1\right> = \left<y_N\right> = \frac{1}{Z}\sum_{i}a_i b_i
\lambda_i^{N-1}
\end{equation}
and

\begin{equation}\label{14}
\left<y_n\right> = \frac{1}{Z}\sum_{i,j}a_i a_j Y_{ij}^{(1)}
\lambda_i^{n-1} \lambda_j^{N-n}
\end{equation}
for $n\neq 1,N$. By evaluating the geometric summation that appears
in Eq. (\ref{9}) when replacing the $\left<y_n\right>$ by their
expressions in Eqs. (\ref{13})-(\ref{14}), we finally get

\begin{equation}\label{15}
\left<y\right> = \frac{2}{ZN}\sum_{i}a_i b_i
\lambda_i^{N-1}+\frac{1}{Z\hspace{0.5 mm} N}\sum_{i,j}a_i a_j
Y_{ij}^{(1)} \lambda_i^{-1}
\lambda_j^{N}\frac{r_{ij}^2-r_{ij}^N}{1-r_{ij}}
\end{equation}
with $r_{ij}=\lambda_i/\lambda_j$. The computation of the
correlation length $\xi$ proceeds along similar lines although it is
more elaborate. A sketch of the derivation and the analytical result
can be found in Appendix A.

\section{Finite-size effects near the critical point}

\subsection{Rounding of the melting transition of DNA}

It is well known that a finite-size system does not exhibit any
phase transition. At the critical point its free energy is analytic
and consequently all thermodynamical quantities are regular. Let $L$
be the size of a system having a critical behaviour in the
thermodynamical limit $L \rightarrow\infty$. For this system
finite-size effects manifest themselves as $e^{-L/\xi}$, where $\xi$
is the correlation length, by rounding the critical point
singularity. In other words they become important over a region for
which $\xi \sim L$. A simple example of this rounding phenomena
concerning the Ising model can be found in \cite{8}. For an
infinite-size Ising system, as the magnetic field $H$ varies, the
order parameter jumps discontinuously from $-M_{cr}$ to $+M_{cr}$ at
the critical point $H=0$. On the other hand if the system's size is
finite, this transition occurs on a finite region of order $\Delta
H\simeq k_B T/(M_{cr}L^d)$ with a large but finite slope $\sim M_L^2
L^d/(k_BT)$ where $M_L$ is the most probable value of the
magnetization in the finite system.

For the two DNA models sketched in Sect. 2, the size $L$ of the
system is equal to the number $N$ of base pairs in the sequence
times the distance between two successive base pairs. This latter
quantity playing no role in the dynamics of the investigated models,
we will henceforth use indistinctly $N$ or $L$ to refer to the size
of the sequence.

Given a sequence of length $N$, the first task consists in
determining its critical temperature, which we denote by $T_c(N)$.
Among the several methods listed for example in \cite{anan}, we
found it rather simple and convenient to search for the maximum of
the specific heat $C_V$, which is more pronounced than that of the
correlation length $\xi$, thus allowing for a more accurate
localization of the temperature. Two observations confirm a
posteriori that the critical temperatures thus obtained are correct.
First, the shift in critical temperature is found to vary as a power
of $N$, as predicted by finite-size scaling theory (see below). The
top plot of Fig. 1 shows for example that $T_c-T_c(N)$, where $T_c$
stands for $T_c(\infty)$, varies as $N^{-1.00}$ and $N^{-1.05}$ for
the DPB and JB models, respectively. Note that this scaling is also
in excellent agreement with the semi-empirical formula used by
experimentalists to calculate the melting temperature of finite
sequences. Moreover, as will be seen later (Figs. 3 and 4) the
curves for the temperature evolution of $C_V$, $\xi$,
$\left<y\right>$, etc... for sequences with different lengths $N$
all coincide sufficiently far from the critical temperature when
plotted as a function of the reduced temperature

\[t(N)=\frac{T-T_c(N)}{T_c(N)}\hspace{0.5 mm}.\]
In order to illustrate finite-size effects acting on DNA melting, we
first computed the entropy per base, $s=S/N$, for an infinite chain
and a short DNA sequence for both the DPB and the JB models. Results
are shown in Fig. 2. At the thermodynamic limit, the entropy $s$ is
clearly discontinuous at the critical temperature, as is expected
for first order phase transitions. In contrast, smooth curves are
observed over the whole temperature range for the sequence with
$N=100$. We next computed the specific heat per base, $c_V=C_V/N$,
for increasing sequence length and temperature. The top and bottom
plots of Fig. 3 show the temperature evolution of $c_V$ for seven
values of $N$ ranging from 100 to infinity for the DPB and JB
models, respectively. It is seen in this figure that rounding
manifests itself through a decrease in the maximum of $c_V$ as $N$
decreases, but also through the fact that the sharp rise of $c_V$
takes place further and further from the critical temperature, that
is, at increasingly larger values of  $|t(N)|$. This is particularly
clear for the DPB model, which at the thermodynamic limit undergoes
a very sharp transition, i.e. a transition that is noticeable only
at very small values of $|t|=|t(\infty)|$ \cite{21}. Quite
interestingly, examination of Fig. 3 also indicates that the two
models consequently give very comparable results up to $N\simeq
1000$, while the narrower nature of the phase transition for the DPB
model becomes apparent for longer sequences.

At this point, it should be emphasized that boundary effects may
become important when the size of the system is small. In order to
check whether such boundary effects play a role in the results
presented above, we repeated these computations with periodic
boundary conditions instead of open ones and found that this alters
only very little the results for $s$ and $c_V$ down to $N=100$.
Conclusion is therefore that boundary effects play only a marginal
role down to this size.

Finally, we computed, for the JB model, the temperature evolution of
the correlation length $\xi$ (Eq. (\ref{17})) and the order
parameter $\left<y\right>$ (Eq. (\ref{15})) for increasing values of
$N$. As the critical temperature is approached, the correlation
length $\xi$ of a finite-size system is expected to increase
according to the power law $\xi\propto t^\nu(N)$ till it reaches the
system's dimension $L$ and freezes. This behaviour can be checked in
the middle plot of Fig. 1, which shows the evolution of the maximum
of the correlation length (in units of the separation between
successive base pairs) as a function of $N$ : the maximum of $\xi$
is indeed of the same order of magnitude as $N$ and the curve scales
as $N^{0.97}$. An exception however occurs for the last three points
with $N\geq 3000$ : we will come back later to this point. The
bottom plot of Fig. 4 additionally shows the temperature evolution
of $\xi$ for seven values of $N$ ranging from 100 to infinity. One
observes just the same rounding effects as for the specific heat in
Fig. 3. This is again the case for the temperature evolution of the
order parameter $\left<y\right>$, which is drawn in the top plot of
Fig. 4. This latter plot however displays a remarkable feature, in
the sense that all the curves converge to the same limit at
$T_c(N)$. To understand why this is the case, it must be realized
that  $\left<y\right>$, the average separation between paired bases,
is the only quantity which diverges at the critical temperature
whatever the size $N$ of the sequence, while, for example, $c_V$ and
$\xi$ diverge for infinitely long chains but remain finite for
finite chains. The limit towards which all curves converge in the
top plot of Fig. 4 is thus just the approximation of infinity
imposed by the numerical procedure (size of the grid, etc...).

\subsection{Finite-size scaling analysis}

The basic idea of finite-size scaling is that the correlation length
$\xi$ is the only length that matters close to the critical
temperature and that one just needs to compare the linear dimension
$L$ of the system to  $\xi$  : rounding and shifting indeed set in
as soon as $L/\xi\sim 1$ . Since, by definition of the critical
exponent $\nu$, $\xi$ grows as $t^{-\nu}$, one has
$(L/\xi)\propto{(tL^{1/\nu})}^\nu$ as $L\to\infty$ or, $L$ being
proportional to the number $N$ of paired bases,
$(L/\xi)\propto{(tN^{1/\nu})}^\nu$. In the absence of external
field, it is therefore natural to write the singular part of the
free energy of the finite-size system in the form
\begin{equation}\label{15.1}
f_{sing}=N^{-d}\hspace{0.5mm}Y(tN^{1/\nu})\hspace{0.5mm}.
\end{equation}
where $Y$ is some homogeneous function. Differentiating Eq.
(\ref{15.1}) twice with respect to $t$, one obtains that $c_V$  is
equal to
\begin{equation}\label{15.2}
c_V=N^\rho G(tN^\sigma)\hspace{0.5mm},
\end{equation}
where

\begin{equation}\label{15.3}
\begin{split}
\rho&=\frac{2}{\nu}-d\\
\sigma&=\frac{1}{\nu}
\end{split}
\end{equation}
and $G$ is an homogeneous function which is proportional to the
second derivative of $Y$. By using Josephson's identity,
$2-\alpha=\nu d$, where $\alpha$ is the critical exponent for $c_V$
($c_V\propto t^{-\alpha})$, coefficients $\rho$ and $\sigma$ can be
recast in the form

\begin{equation}\label{15.4}
\begin{split}
\rho&=\frac{\alpha}{\nu}\\
\sigma&=\frac{1}{\nu}
\end{split}
\end{equation}
Conversely, if there occur several lengths that diverge at the
critical point, as is the case for DNA melting, then it is no longer
so clear to which of these lengths $L$ should be compared. In order
to tackle this more complex case, Binder et al \cite{24} derived a
method which is based on the use of an irrelevant variable $u$ and
an expression of the form

\begin{equation}\label{15.5}
f_{sing}=N^{-d}F(tN^{y_t},uN^{y_u})
\end{equation}
After several approximations and a little bit of algebra, these
authors obtain $c_V$ in the form of Eq. (\ref{15.2}) with, however,

\begin{equation}\label{15.6}
\begin{split}
\rho&=\frac{2d}{2\beta+\gamma}-d\\
\sigma&=\frac{2d}{2\beta+\gamma}\hspace{0.5mm}.
\end{split}
\end{equation}

Finally, the lengths that diverge at DNA melting are $\xi$ and
$\left<y\right>$, the average separation between paired bases. One
might wonder whether $L$ should not be compared to $\left<y\right>$
instead of $\xi$. In order to check this hypothesis, let us denote
by $\lambda$ the characteristic exponent for $\left<y\right>$
($\left<y\right>\propto t^\lambda$). Remember that if the external
field is proportional to $y$ then $\left<y\right>$ is the order
parameter $m$, so that $\lambda$ is equal to $\beta$, the critical
exponent for $m$ ($m\propto t^\beta$). Let us next express the
singular part of the free energy in the form

\begin{equation}\label{15.7}
f_{sing}=N^{-d}F\left(tN^{-1/\lambda}\right)\hspace{0.5mm}.
\end{equation}
Differentiating Eq. (\ref{15.7}) twice with respect to $t$, one
again obtains $c_V$ in the form of Eq. (\ref{15.2}) with, however

\begin{equation}\label{15.8}
\begin{split}
\rho&=-\frac{2}{\lambda}-d\\
\sigma&=-\frac{1}{\lambda}\hspace{0.5mm}.
\end{split}
\end{equation}

Table I shows the values of $\rho$ and $\sigma$ calculated from the
characteristic exponents reported in \cite{21} and Eqs (\ref{15.3}),
(\ref{15.4}), (\ref{15.6}) and (\ref{15.8}), as well as adjusted
values. These latter ones were obtained by varying $\rho$ and
$\sigma$ by hand in order that the plots of $c_V/N^\rho$ as a
function of $tN^\sigma$ are superposed for an interval of values of
$N$ as large as possible. By setting $t=0$ in Eq. (\ref{15.2}), one
sees that the maximum of $c_V$ scales as $N^\rho$. $\rho$ was
therefore adjusted in the neighbourhood of the slope of the plot of
the maximum of $c_V$ as a function of $N$ (bottom plot of Fig. 1).
On the other hand, $\sigma$  was adjusted in the neighbourhood of
$1/\nu$. Examination of Table I indicates that the values of $\rho$
and $\sigma$ obtained from Eqs. (\ref{15.3}) and (\ref{15.8})
compare well with the adjusted ones, while this is certainly not the
case for the values obtained from Eqs. (\ref{15.4}) and
(\ref{15.6}). Figs. 5 and 6 further show plots of $c_V/N^\rho$ as a
function of $tN^\sigma$ for, respectively, the DPB and JB models,
and values of $\rho$ and $\sigma$ obtained from Eq. (\ref{15.4})
(top plots) and adjusted ones (bottom plots). It is seen in the top
plots that the curves with different values of $N$ are far from
being superposed for the values of $\rho$ and $\sigma$ obtained from
Eq. (\ref{15.4}), and the situation is still worse with Eq.
(\ref{15.6}). In contrast, the various curves are fairly well
superposed for the adjusted values of $\rho$ and $\sigma$ (see
bottom plots of Figs 5 and 6), as well as those obtained with Eqs.
(\ref{15.3}) and (\ref{15.8}). An exception occurs for the curves
corresponding to the largest values of $N$ in the JB model (bottom
plot of Fig. 6). Remember that the corresponding points also depart
from the power law in the bottom plot of Fig. 1. The reason for this
is that the TI method fails to give correct values of
thermodynamical observables too close to the phase transition
discontinuity. Examination of the top plots of Fig. 3 shows that
sequences of length $N=10000$ are still rather far from the
thermodynamic limit for the DPB model, so that one needs not to
worry about the effect of the discontinuity on TI calculations. In
contrast, the bottom plot of Fig. 3 and the two plots of Fig. 4
indicate that sequences of length $N=10000$ have reached the
thermodynamic limit for the JB model, so that the perturbative
effect of the discontinuity becomes noticeable in TI calculations.

The fact that Eq. (\ref{15.3}) leads to a correct superposition of
the curves for different values of the sequence length $N$ is the
proof that the basic hypothesis of finite-size scaling theory is
satisfied. Since, however, Eq. (\ref{15.8}) also leads to a correct
superposition of the curves, it is, as expected, no longer clear to
which diverging length ($\xi$ or $\left<y\right>$) $L$ should be
compared, both possibilities leading to a reasonable result. On the
other hand, the fact that curves with different $N$ are no longer
superposed when Eq. (\ref{15.4}) is used to calculate $\rho$ and
$\sigma$ simply reflects the fact that Josephson's identity,
$2-\alpha=\nu d$, is not valid for these two models of DNA melting,
a conclusion which was already arrived at in our preceding work
\cite{21}. Finally, the fact that curves also do not superpose when
Eq. (\ref{15.6}) is used indicates that one of the several
hypotheses made by the authors of Ref. \cite{24} to arrive to these
expressions is not satisfied for the DNA models, although it is not
an easy task to tell which one(s) is(are) invalidated.
Alternatively, Eq. (\ref{15.6}) can be straightforwardly derived
from Eq. (\ref{15.3}) by using Rushbrooke identity
($\alpha+2\beta+\gamma=2$) as well as Josephson's one. This latter
identity being not valid, it comes as no surprise that Eq.
(\ref{15.6}) leads to as poor a result as Eq. (\ref{15.4}).

\section{Conclusion}

To summarize, we modified the Transfer Integral method to adapt it
to the calculation of thermodynamic quantities of finite sequences
with open boundary conditions. Non-extensive quantities, like the
average separation of paired bases $\left<y\right>$  and the
correlation length $\xi$, turned out to be the most tricky ones to
evaluate. We then applied this modified procedure to the DPB and JB
dynamical models, in order to clarify how the finiteness of the
length of the sequence affects the phase transition that takes place
at DNA melting temperature. We showed that the rounding of the
transition that occurs when the size of the sequence decreases is
clearly reflected in the temperature evolution of most quantities,
including the specific heat  $c_V$, the correlation length   $\xi$,
and the average separation of paired bases  $\left<y\right>$. We
next performed a finite-size scaling analysis of the two systems and
showed that the singular part of the free energy can indeed be
expressed in terms of an homogeneous function. However, since both
$\xi$ and $\left<y\right>$ diverge at the melting transition, it is
no longer clear whether the argument of the homogeneous function
should be (a power of) $L/\xi$ or $L/\left<y\right>$. Moreover,
Josephson's identity is satisfied for none of the investigated
systems, so that the derivation of the characteristic exponents
$\rho$ and $\sigma$, which appear in the asymptotic expression of
the specific heat $c_V$, requires some care.

The Transfer Integral (TI) method appears as the only efficient
numerical tool to study the thermodynamics of DNA melting in detail.
In the formulation used here, it however applies only to homogeneous
chains, while it is well established that the heterogeneity of real
DNA molecules may smear out the discontinuity of the melting
transition, just like the finiteness of the sequence does. Our next
goal is therefore to overcome the technical difficulty associated
with the application of the TI method to inhomogeneous chains and
investigate the effect of heterogeneities on the phase transition at
DNA melting.

\appendix
\section{Computation of the correlation length}

The \textit{static form factor} is defined as

\begin{equation}\label{16}
S(q) = \left<\left|\sum_{n=1}^N
(y_n-\left<y_n\right>)e^{iqan}\right|^2\right>
\end{equation}
and the correlation length is given by

\begin{equation}\label{17}
\xi^2=-\left.\frac{1}{2S(q)}\frac{d^2S(q)}{dq^2}\right|_{q=0}.
\end{equation}
We stress that the statistical weight of the bases at the
extremities $n=1,N$ is different from that of the other ones, so
that they must be treated separately. We first write Eq. (\ref{16})
in the explicit form

\begin{equation}\label{18}
S(q)=\sum_{n=1}^N \sum_{m=1}^N \left<\delta y_n \delta
y_m\right>e^{iqa(n-m)}
\end{equation}
where $\delta y_n=y_n-\left<y_n\right>$. By isolating averages
concerning extremity values, we get

\begin{equation}\label{19}
\begin{split}
S(q)&=\left<\delta y_1^2\right>+\left<\delta y_N^2\right>+
2\left<\delta y_1\delta y_N\right> \cos[qa(N-1)]\\
&+S_1 e^{iqa}+S_1^*e^{-iqa}+S_N e^{iNqa}+S_N^*e^{-iNqa}+S_{mid}
\end{split}
\end{equation}
where

\begin{equation}\label{20}
\begin{split}
S_1&=\sum_{m=2}^{N-1}\left<\delta y_1\delta y_m\right>e^{-iqam}\\
S_N&=\sum_{m=2}^{N-1}\left<\delta y_m\delta y_N\right>e^{-iqam}\\
S_{mid}&=\sum_{m=2}^{N-1}\sum_{n=2}^{N-1}\left<\delta y_n\delta
y_m\right>e^{iqa(n-m)}.
\end{split}
\end{equation}
Defining

\begin{equation}\label{21}
\begin{split}
c_i &= \int \,dy e^{-\beta V_M(y)/2} y^2\Phi_i(y)\\ Y_{ij}^{(2)} &=
\int \,dy\Phi_i(y)y^2\Phi_j(y)\hspace{1 mm}.
\end{split}
\end{equation}
we obtain the relations :

\begin{equation}\label{22}
\begin{split}
\left<y_1^2\right> &= \left<y_N^2\right> = \frac{1}{Z}\sum_{i}a_i
c_i \lambda_i^{N-1}\\
\left<y_1y_N\right> &= \frac{1}{Z}\sum_{i}b_i^2 \lambda_i^{N-1}\\
\left<y_1y_m\right> &= \frac{1}{Z}\sum_{ij}a_j\hspace{0.3 mm}
b_i\lambda_i^{m-1}\lambda_j^{N-m} Y_{ij}^{(1)}\\
\left<y_my_N\right> &= \frac{1}{Z}\sum_{ij}a_i\hspace{0.3 mm}b_j
\lambda_i^{m-1}\lambda_j^{N-m} Y_{ij}^{(1)}\\
\left<y_n^2\right> &= \frac{1}{Z}\sum_{ij}a_i\hspace{0.3 mm}
a_j\lambda_i^{n-1}\lambda_j^{N-n} Y_{ij}^{(2)}\\
\left<y_n y_m\right> &= \frac{1}{Z}\sum_{ijk}a_i a_k
Y_{ij}^{(1)}Y_{jk}^{(1)}\cdot \left\{\begin{array}{ccc}
\lambda_i^{n-1}\lambda_j^{m-n}\lambda_k^{N-m}
\hspace{1 cm}\mbox{if}\hspace{2 mm}m>n\hspace{2 mm}\\
\lambda_i^{m-1}\lambda_j^{n-m}\lambda_k^{N-n} \hspace{1
cm}\mbox{if}\hspace{2 mm}m<n\hspace{1 mm}.
\end{array}\right.
\end{split}
\end{equation}
According to the relations in Eqs. (\ref{14}) and (\ref{22}), the
summations in Eq. (\ref{18}) are just geometric series, which we
evaluated formally in order to increase the speed of numerical
calculations by a factor $N^2$. After some tedious algebra, one
obtains :
\[S(q) = \left<\delta y_1^2\right>+\left<\delta
y_N^2\right>+2\left<\delta y_1\delta
y_N\right>\cos[qa(N-1)]+\sum_{ij}H_{ij}\frac{f_{ij}(2)-f_{ij}(N)}{1-f_{ij}(1)}\hspace{8
cm}\]
\[+\sum_{i,j}\frac{1}{\cosh(\alpha_{ij})-\cos(qa)}\left\{[D_{ij}f_{ij}(1)+C_{ij}f_{ij}(N)]\cos[(N-2)qa]
-D_{ij}f_{ij}(N-1)-C_{ij}f_{ij}(2)\right.\hspace{4.1 cm}\]
\[\left.-[D_{ij}f_{ij}(2)+C_{ij}f_{ij}(N-1)]\cos[(N-1)qa]+[D_{ij}f_{ij}(N)+C_{ij}f_{ij}(1)]\cos[qa]\right\}\]
\begin{equation}\label{23}
+\sum_{ijk}\frac{2M_{ijk}}{(1-g(1,1))(1+f_{ij}(2)-2f_{ij}(1)\cos(qa))(1+f_{jk}(2)-2f_{jk}(1)\cos(qa))}\hspace{5
cm}
\end{equation}
\[\times\left\{-g(3,3)-g(4,4)-g(2,4)+g(N,N)+g(N,N+2)+g(N+1,N+1)\right.\hspace{5 cm}\]
\[+[g(2,3)-g(N,N+1)+2g(3,4)-g(N+1,N+2)+g(4,3)-g(N+1,N)-g(N,N+1)]\cos(qa)\]
\[+[g(N+1,N+1)-g(3,3)]\cos(2qa)+[g(2,N+1)-g(3,N+2)]\cos[(N-3)qa]\hspace{4 cm}\]
\[\left.+[g(4,N+2)-g(2,N)]\cos[(N-2)qa]+[g(3,N)-g(4,N+1)]\cos[(N-1)qa]\right\}\hspace{3 cm}\]
\[+\left|\sum_{ij}M_{ij}\frac{e^{2(iqa-\alpha_{i,j})}-e^{N(iqa-\alpha_{ij})}}{1-e^{iqa-\alpha_{ij}}}\right|^2,\hspace{12 cm}\]
where

\begin{equation}\label{24}
\begin{split}
\alpha_{ij}&=-\ln(r_{ij})\\
M_{ij}&=\frac{1}{Z}\lambda_i^{-1}\lambda_j^{N}a_ia_jY^{(1)}_{ij}\\
H_{ij}&=\frac{1}{Z}\lambda_i^{-1}\lambda_j^{N}a_ia_jY^{(2)}_{ij}\\
T_{ij}&=\frac{1}{Z}\lambda_i^{-1}\lambda_j^{N}b_ia_jY^{(1)}_{ij}\\
G_{ij}&=\frac{1}{Z}\lambda_i^{-1}\lambda_j^{N}a_ib_jY^{(1)}_{ij}\\
C_{ij}&=T_{ij}-\left<y_1\right>M_{ij}\\
D_{ij}&=G_{ij}-\left<y_1\right>M_{ij}\\
f_{ij}(n)&=e^{-n\alpha_{ij}}\\
g(n,m)&=e^{-n\alpha_{ij}-m\alpha_{jk}}\hspace{1 mm}.
\end{split}
\end{equation}

Finally, $S(q)$ must be derivated twice with respect to the wave
vector $q$ in order to get the correlation length from Eq.
(\ref{17}).

\newpage
\vspace{1cm}\hspace{6cm} \textbf{TABLE CAPTION}

\vspace{1cm}
\textbf{Table I :} Values of the coefficients $\rho$
and $\sigma$ of Eq. (\ref{15.2}) for the DPB and JB models. First
four lines show the values calculated from the characteristic
exponents reported in Ref. \cite{21} and Eqs (\ref{15.3}),
(\ref{15.4}), (\ref{15.6}) and (\ref{15.8}). The last line shows the
values adjusted by hand in order that the plots of $c_V/N^\rho$ as a
function of $tN^\sigma$ are superposed on an interval of values of
$N$ as large as possible (see bottom plots of Figs. 5 and 6).

\newpage
\vspace{1cm}\hspace{6cm}\textbf{FIGURE CAPTIONS}

\vspace{1cm} \textbf{Figure 1 :} (color online) Log-log plots, as a
function of the sequence length $N$, of the reduced critical
temperature shift $1-T_c(N)/T_c(\infty)$ (top plot), the maximum of
the correlation length $\xi$ (middle plot), and the maximum of $c_V$
(bottom plot), according to the DPB (squares) and JB (circles)
models. $\xi$ is in units of the separation between two successive
base pairs, and $c_V$ in units of $k_B$. The solid and dash-dotted
lines show the result of the adjustment of power laws against the
calculated points.

\vspace{5mm} \textbf{Figure 2 :} (color online) Plot of the entropy
per site $s$ as a function of the rescaled temperature $t(N)$ for an
infinitely long chain (circles) and a sequence with $N=100$ bp
(squares), according to the DPB model (top plot) and the JB one
(bottom plot). $s$ is in units of $k_B$ .

\vspace{5mm}\textbf{Figure 3 :} (color online) Log-log plots of the
specific heat per site $c_V$ as a function of the opposite $-t(N)$
of the rescaled temperature for the DPB model (top plot) and the JB
one (bottom plot) and seven values of the sequence length $N$
ranging from $100$ to $\infty$. $c_V$ is in units of $k_B$. Note
that, at the thermodynamic limit of infinitely long chains, $c_V$
becomes infinite at the critical temperature but numerical
limitations of the TI method prevent the observation of such
divergence.

\vspace{5mm}\textbf{Figure 4 :} (color online) Log-log plots of the
correlation length $\xi$ (bottom plot) and the order parameter
$\left<y\right>$ (top plot) as a function of the opposite $-t(N)$ of
the rescaled temperature for the JB model and seven values of the
sequence length $N$ ranging from $100$ to $\infty$. $\xi$ is in
units of the separation between two successive base pairs and
$\left<y\right>$ in units of the inverse $1/a$ of the Morse
potential parameter. Although numerical limitations of the TI method
prevent the observation of such divergences, $\xi$ becomes infinite
at the critical temperature at the thermodynamic limit of infinitely
long chains, while $\left<y\right>$ becomes infinite at the critical
temperature whatever the length $N$ of the sequence.

\vspace{5mm}\textbf{Figure 5 :} (color online) Plots, for six values
of the sequence length $N$ ranging from $100$ to $10000$, of
$c_V/N^\rho$ as a function of $tN^\sigma$ for the DPB model and
values of $\rho$ and $\sigma$ obtained from Eq. (\ref{15.4}) (top
plot) or adjusted by hand (bottom plot).

\vspace{5mm}\textbf{Figure 6 :} (color online) Plots, for six values
of the sequence length $N$ ranging from $100$ to $10000$, of
$c_V/N^\rho$ as a function of $tN^\sigma$ for the JB model and
values of $\rho$ and $\sigma$ obtained from Eq. (\ref{15.4}) (top
plot) or adjusted by hand (bottom plot).

\newpage
\begin{table}[htbp]
\begin{tabular}{ccccc}
\hline\hline
\multicolumn{5}{c}{\hspace{4.0cm}DPB model\hspace{3.6cm}JB model}\\
&\hspace{2cm}$\rho$&\hspace{2cm}$\sigma$&\hspace{2cm}$\rho$&\hspace{2cm}$\sigma$\\
\hline
Eq. (\ref{15.3})&\hspace{2cm}0.79&\hspace{2cm}0.89&\hspace{2cm}0.63&\hspace{2cm}0.81\\
Eq. (\ref{15.4})&\hspace{2cm}1.29&\hspace{2cm}0.89&\hspace{2cm}0.92&\hspace{2cm}0.81\\
Eq. (\ref{15.6})&\hspace{2cm}1.78&\hspace{2cm}1.39&\hspace{2cm}1.82&\hspace{2cm}1.41\\
Eq. (\ref{15.8})&\hspace{2cm}0.87&\hspace{2cm}0.93&\hspace{2cm}0.53&\hspace{2cm}0.76\\
adjusted &\hspace{2cm}0.85&\hspace{2cm}1.00&\hspace{2cm}0.45&\hspace{2cm}0.90\\
\hline\hline
\end{tabular}
\caption{}
\end{table}

\begin{figure}
\includegraphics[width=18cm]{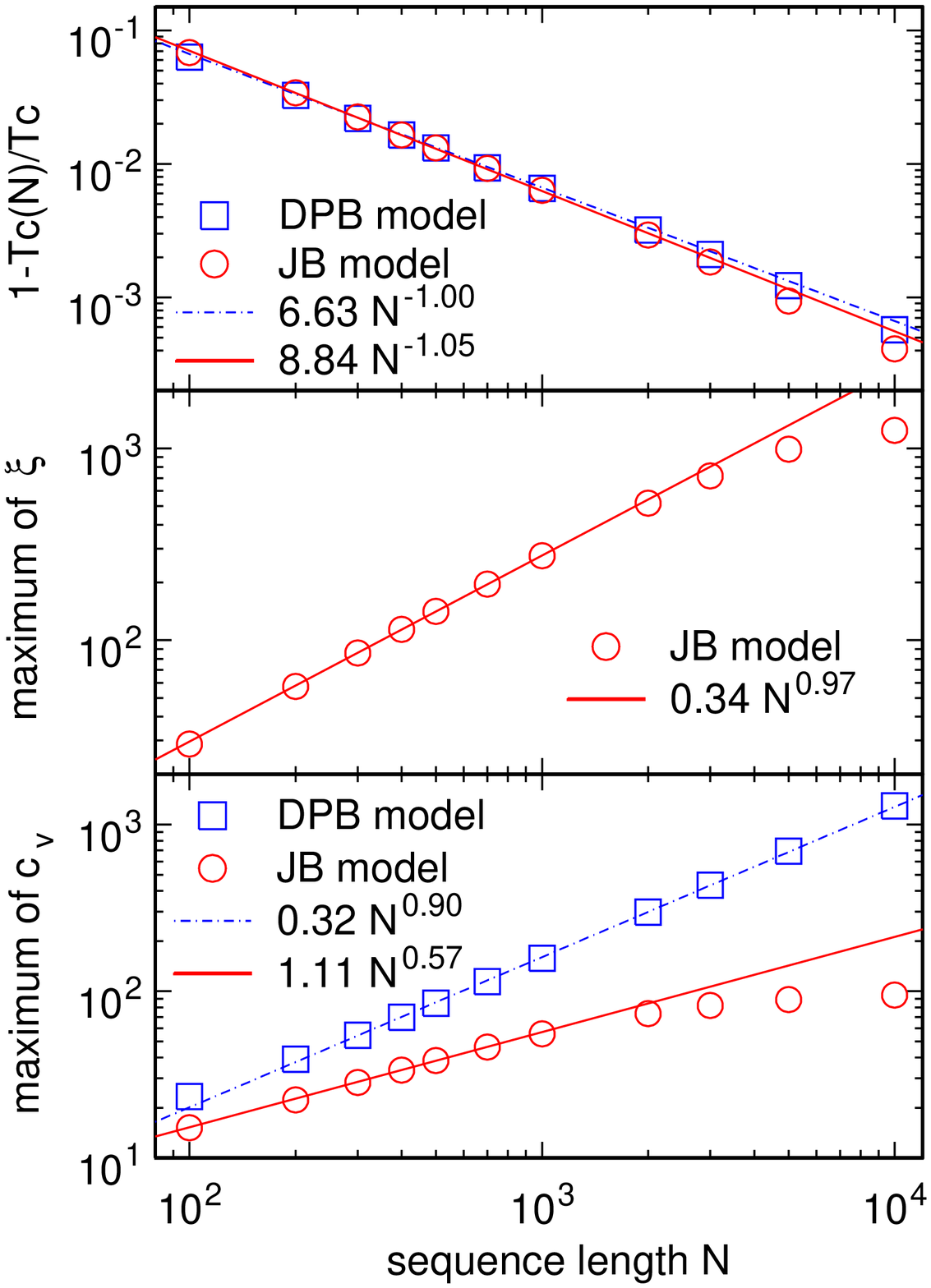}
\caption{\footnotesize}
\end{figure}

\begin{figure}
\includegraphics[width=18cm]{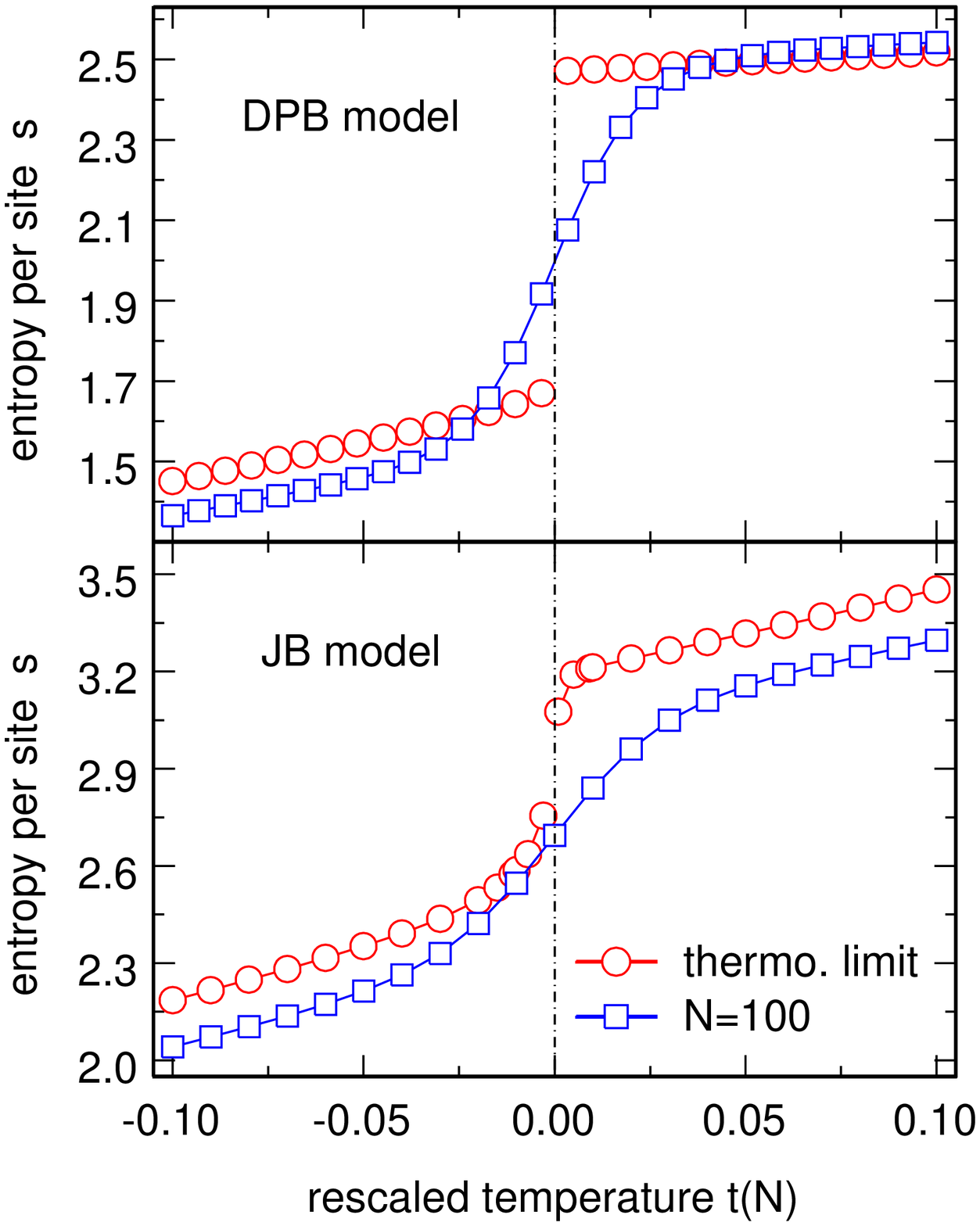}
\caption{\footnotesize}
\end{figure}

\begin{figure}
\includegraphics[width=18cm]{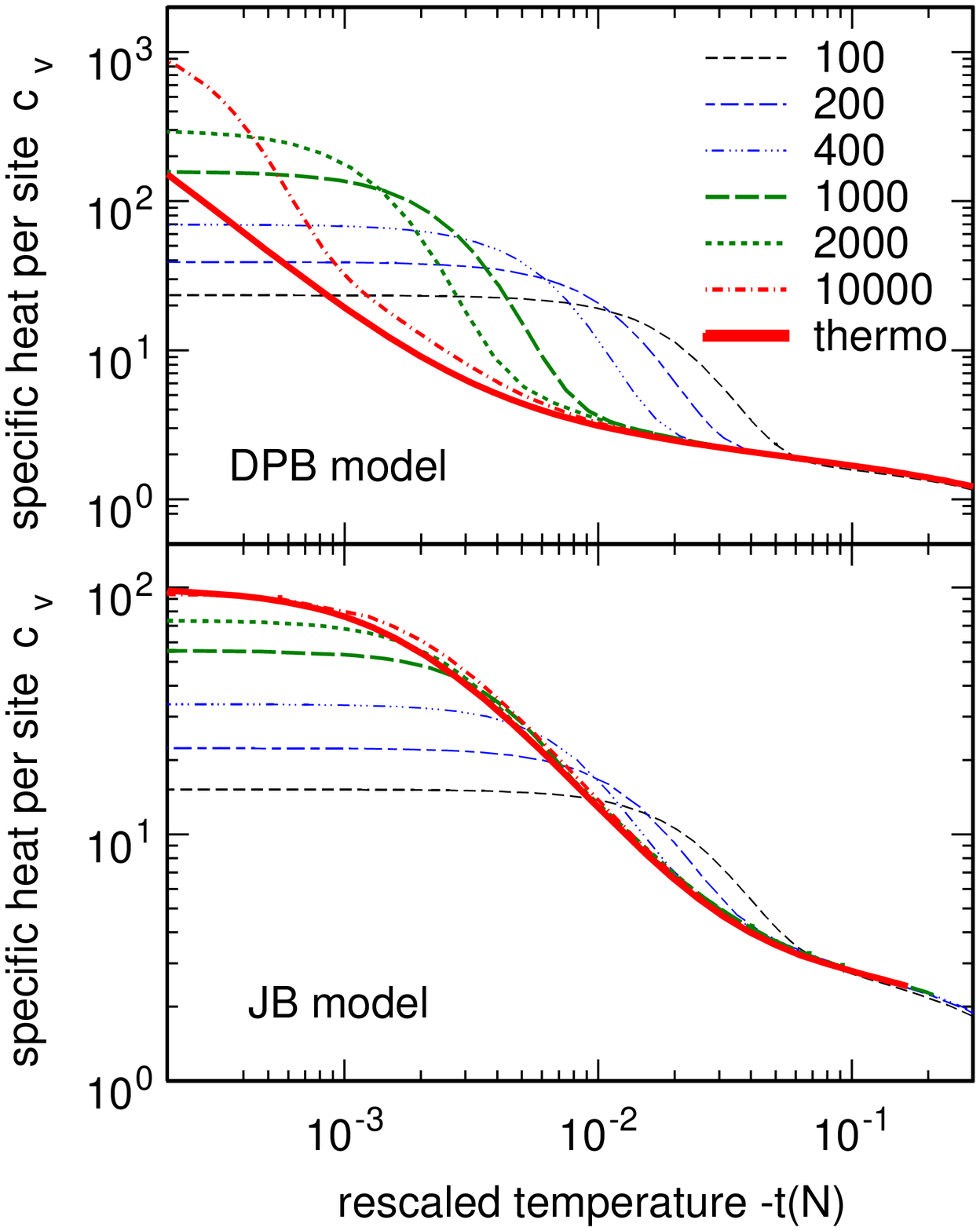}
\caption{\footnotesize}
\end{figure}

\begin{figure}
\includegraphics[width=18cm]{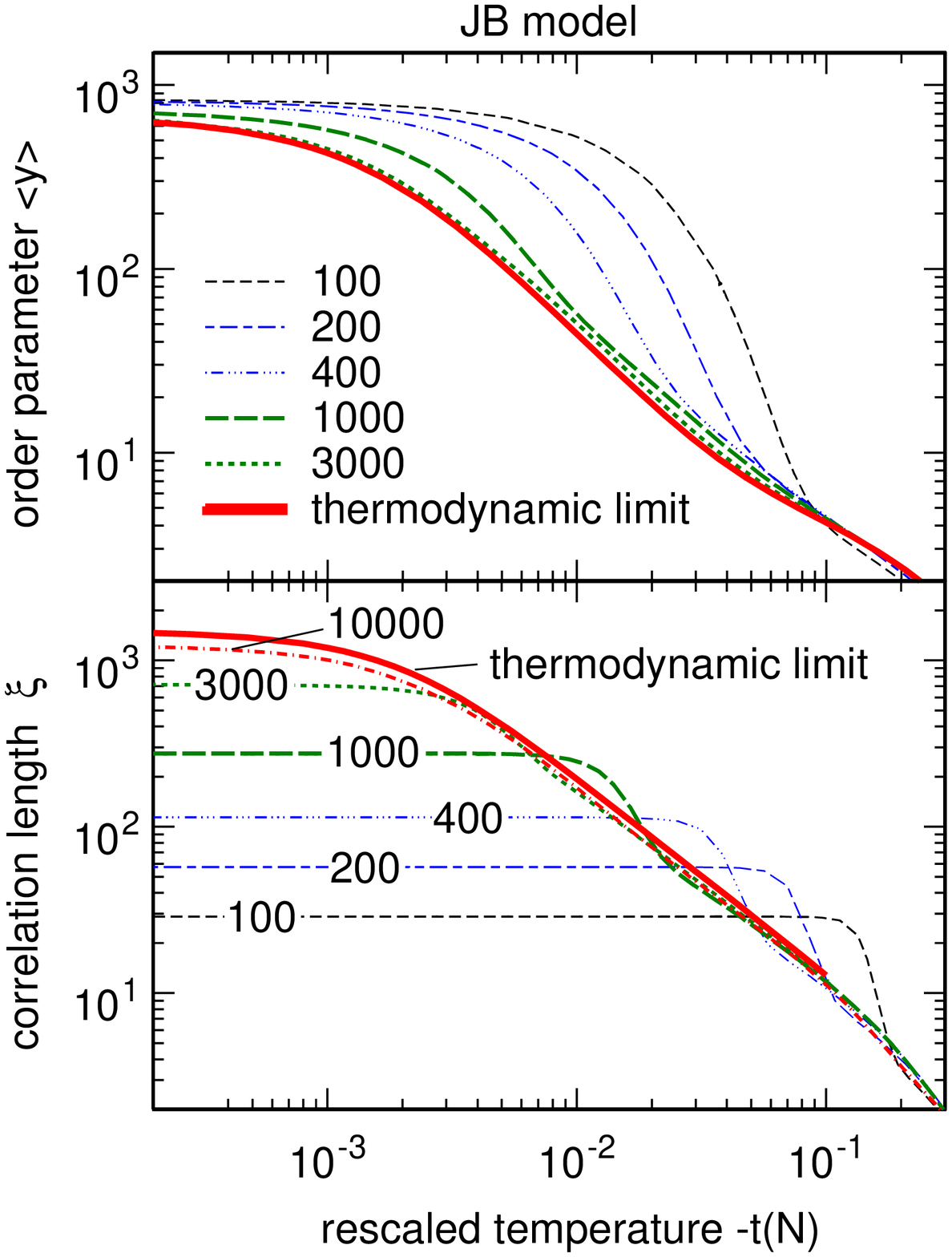}
\caption{\footnotesize}
\end{figure}

\begin{figure}
\includegraphics[width=18cm]{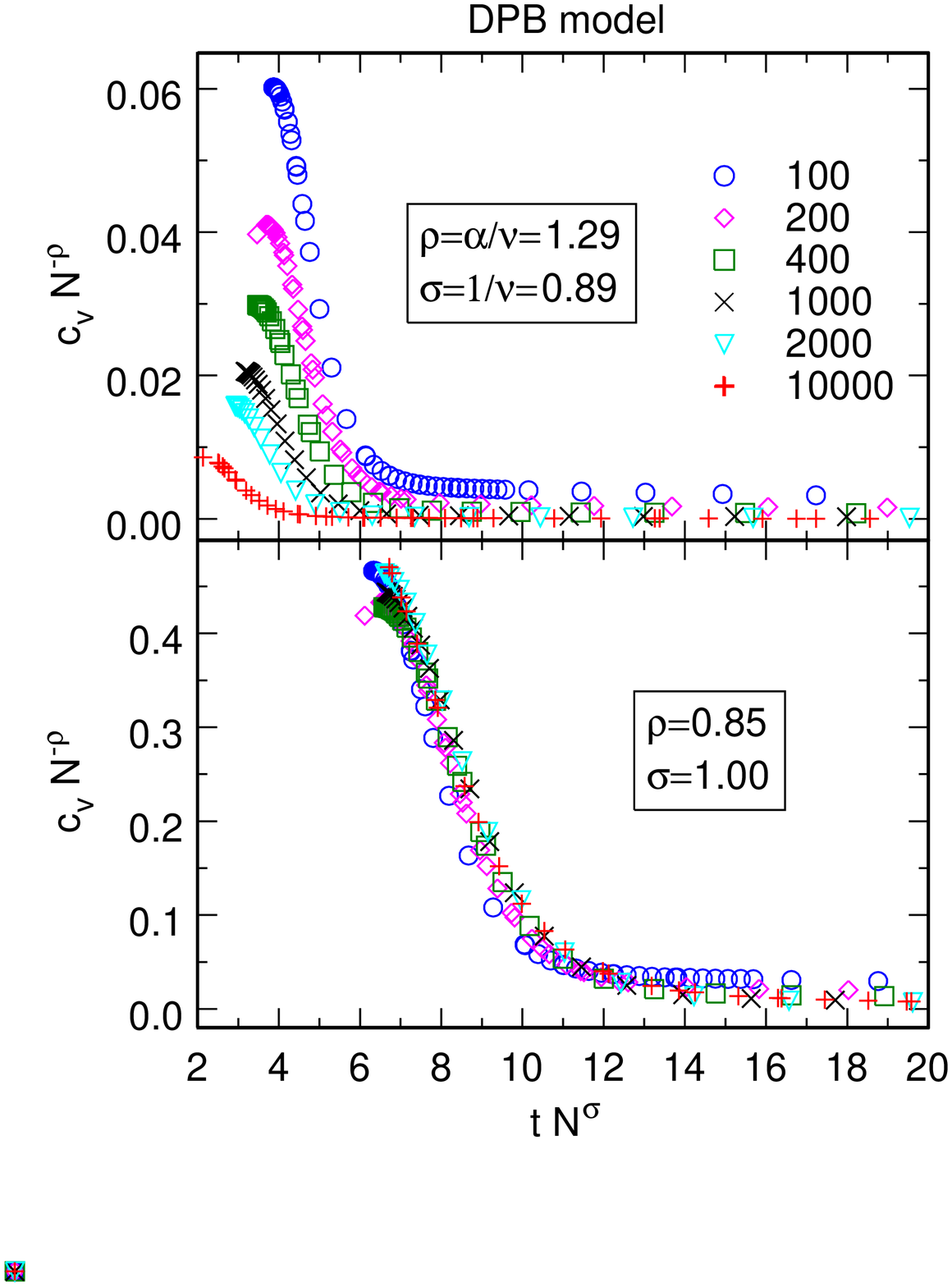}
\caption{\footnotesize}
\end{figure}

\begin{figure}
\includegraphics[width=18cm]{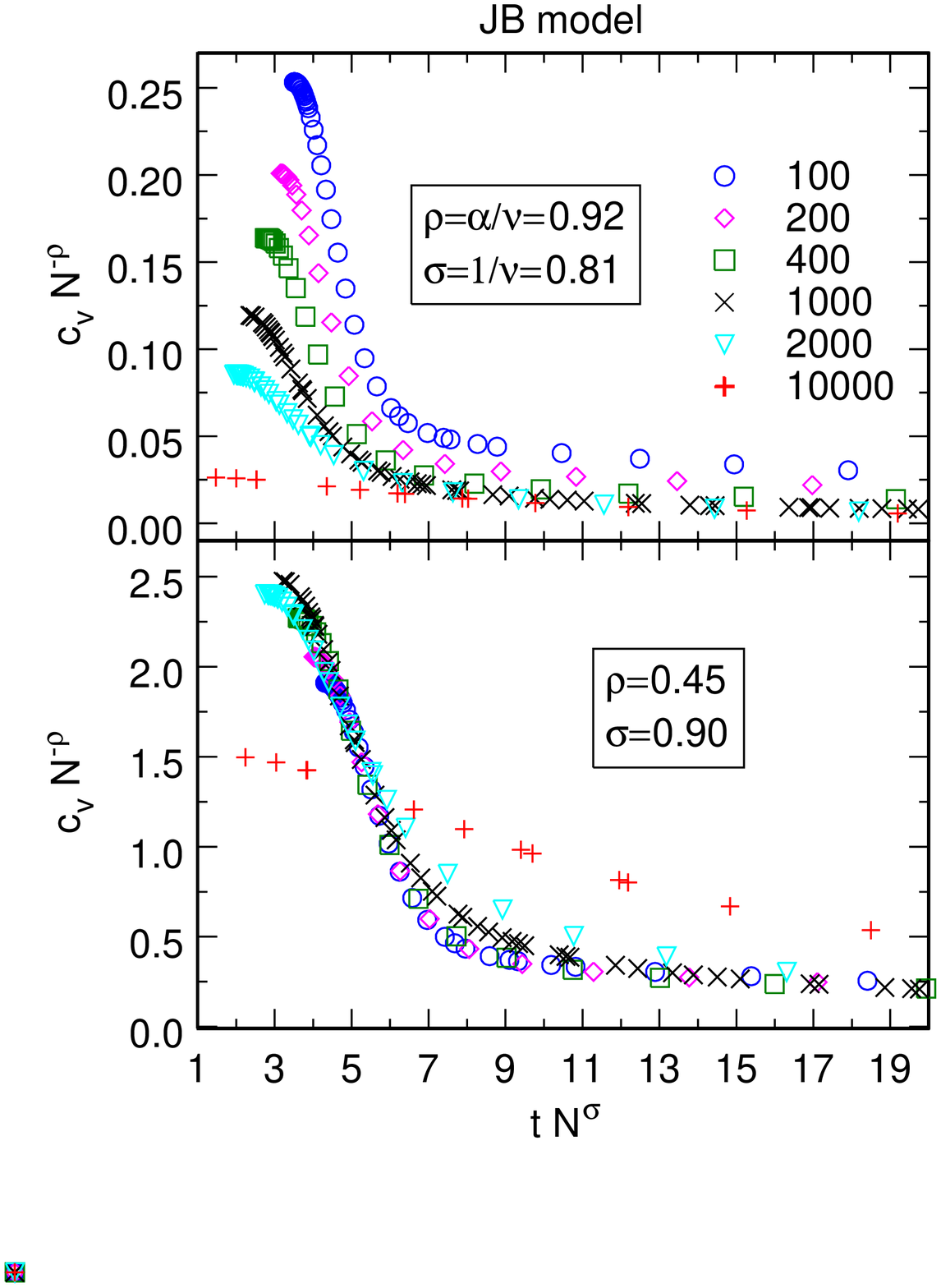}
\caption{\footnotesize}
\end{figure}

\end{document}